\documentclass[floatfix,a4paper,prb,twocolumn,superscriptaddress, showpacs,preprintnumbers,amsmath,amssymb]{revtex4}

\usepackage{hyperref}

\usepackage{graphicx}
\usepackage{color}
\usepackage{multirow, ulem}

\begin{document}

\title{Local electrical tuning of the nonlocal signals in a Cooper pair splitter}

\author{G.~F\"{u}l\"{o}p}
\thanks{These authors contributed equally to this work.}
\affiliation{Department of Physics, Budapest University of Technology and Economics, and Condensed Matter Research Group of the Hungarian Academy of Sciences, Budafoki \'{u}t 8, 1111 Budapest, Hungary}

\author{S.~d'Hollosy}
\thanks{These authors contributed equally to this work.}
\affiliation{Department of Physics, University of Basel, Klingelbergstrasse 82, CH-4056 Basel, Switzerland}

\author{A.~Baumgartner}
\email{andreas.baumgartner@unibas.ch}
\affiliation{Department of Physics, University of Basel, Klingelbergstrasse 82, CH-4056 Basel, Switzerland}

\author{P.~Makk}
\affiliation{Department of Physics, Budapest University of Technology and Economics, and Condensed Matter Research Group of the Hungarian Academy of Sciences, Budafoki \'{u}t 8, 1111 Budapest, Hungary}
\affiliation{Department of Physics, University of Basel, Klingelbergstrasse 82, CH-4056 Basel, Switzerland}

\author{V.A.~Guzenko}
\affiliation{Laboratory for Micro- and Nanotechnology, Paul Scherrer Institute, CH-5232 Villigen PSI, Switzerland}

\author{M.H. Madsen}
\altaffiliation[present address: ]{Danish Fundamental Metrology, DK-2800 Kgs. Lyngby, Denmark}
\affiliation{Center for Quantum Devices and Nano-Science Center, Niels Bohr Institute, University of Copenhagen, Universitetsparken 5, DK-2100
Copenhagen, Denmark}

\author{J.~Nyg{\aa}rd}
\affiliation{Center for Quantum Devices and Nano-Science Center, Niels Bohr Institute, University of Copenhagen, Universitetsparken 5, DK-2100
Copenhagen, Denmark} 

\author{C.~Sch\"{o}nenberger}
\affiliation{Department of Physics, University of Basel, Klingelbergstrasse 82, CH-4056 Basel, Switzerland}

\author{S.~Csonka}
\affiliation{Department of Physics, Budapest University of Technology and Economics, and Condensed Matter Research Group of the Hungarian Academy of Sciences, Budafoki \'{u}t 8, 1111 Budapest, Hungary}

\date{\today}

\begin{abstract}
A Cooper pair splitter consists of a central superconducting contact, S, from which electrons are injected into two parallel, spatially separated quantum dots (QDs). This geometry and electron interactions can lead to correlated electrical currents due to the spatial separation of spin-singlet Cooper pairs from S. We present experiments on such a device with a series of bottom gates, which allows for spatially resolved tuning of the tunnel couplings between the QDs and the electrical contacts and between the QDs. Our main findings are gate-induced transitions between positive conductance correlation in the QDs due to Cooper pair splitting and negative correlations due to QD dynamics. Using a semi-classical rate equation model we show that the experimental findings are consistent with in-situ electrical tuning of the local and nonlocal quantum transport processes. In particular, we illustrate how the competition between Cooper pair splitting and local processes can be optimized in such hybrid nanostructures.
\end{abstract}

% 73.23.-b Electronic transport in mesoscopic systems 
% 73.63.Nm Quantum wires  
% 74.45.+c Proximity effects; Andreev reflection; SN and SNS junctions
% 03.67.Bg Entanglement production and manipulation 

\maketitle

\section{Introduction}

Complex top-down electronic nanostructures with a large number of local gates have become of great interest, e.g., for experiments in gate defined quantum rings\cite{Gustavsson_Ihn_Ensslin_Nanolett_2008} or double-quantum dots\cite{Hu_Churchill_Marcus_NatureNano2_2007} with charge detectors, to study possible Majorana Fermions in semiconducting nanowires,\cite{Mourik_Kouwenhoven_Science_2012} or to form and shape QDs on suspended carbon nanotubes, giving control over the coupling between the electrical and the mechanical degrees of freedom.\cite{Benyamini_Iliani_NaturePhys_2014}

Local control of the device parameters is also essential in a device for Cooper pair splitting (CPS). A CPS device is shown in Figs.~\ref{Figure1}a and \ref{Figure1}b and consists of two quantum dots (QD1 and QD2) connected in parallel to a central superconducting contact, S, and to two individual normal metal contacts N1 and N2. The electrons in a superconductor form spin-singlet Cooper pairs which can be separated (split) coherently into N1 and N2 by the interactions on the QDs, resulting in a stream of spatially separated entangled electron pairs.\cite{Recher_Loss_PRB63_2001, lesovik_martin_blatter_EPJB01, Sauret_2004_PRB70_2004} Experimental evidence for correlated currents have been reported recently for devices based on InAs nanowires (NWs)\cite{Hofstetter2009, Hofstetter_Baumgartner_PRL107_2011, Das_Heiblum_NatureComm_2012} and carbon nanotubes,\cite{Herrmann_Kontos_Strunk_PRL104_2010, Schindele_Baumgartner_Schoenenberger_PRL109_2012, Schindele_Baumgartner_PRB89_2014} with efficiencies up to 90\%.\cite{Schindele_Baumgartner_Schoenenberger_PRL109_2012} The CPS efficiency and the relevant physical processes depend strongly on the tunnel couplings, which were determined in previous devices by poorly controlled details in the fabrication process. 
A maximum CPS efficiency is expected for a reasonably strong tunnel coupling of the QDs to the normal contacts and a weaker coupling to S, and for a large, sharp energy gap in the superconductor.\cite{Recher_Loss_PRB63_2001}

Here we present experiments on an InAs NW CPS device fabricated on top of an array of narrow bottom gates. These gates allow us to form QDs at different positions and to tune individually and in-situ the QD-lead tunnel couplings, the chemical potentials of the QDs and the inter-dot coupling.\cite{Fasth_Fuhrer_Samuelson_Nanolett_2005} The superconducting contact is made of Nb, which has a $\sim15$ times larger bulk energy gap than Al, the superconductor used in most previous experiments. In the presented transport experiments we find that the correlation between the conductances of the two QDs depends strongly on the gate configurations. In particular, we tune the barriers on the normal and the superconductor side of the QDs, as well as the inter-dot coupling between two dots, each inducing a transition from positive to negative correlations. We qualitatively reproduce the experimental findings in a semi-classical rate equation model and attribute the observed transitions in the conductance correlations to the competition between the different transport processes on QDs with a finite average population.

Our results shed light on the electron dynamics in such systems and are fundamental for controlling and maximizing the CPS efficiency, as required to detect electron entanglement by violating Bell's inequality,\cite{Kawabata_JPhysSocJpn70_2001, Chtchelkatchev_Blatter_Lesovik_Martin_PRB66_2002, Samuelsson_Buettiker_PRL91_2003, Bednorz_Belzig_PRB83_2011, Braunecker_Levi_Yeyati_PRL111_2013} by an entanglement witness,\cite{Klobus_Baumgartner_Martinek_PRB89_2014} or by using micro wave photons.\cite{Cottet_PRB86_2012, Scherubl_Csonka_PRB89_2014}

% ------------- Fabrication and characterization ---------------------------
\section{Sample fabrication and characterization}

\begin{figure}[t]
\centering
\includegraphics{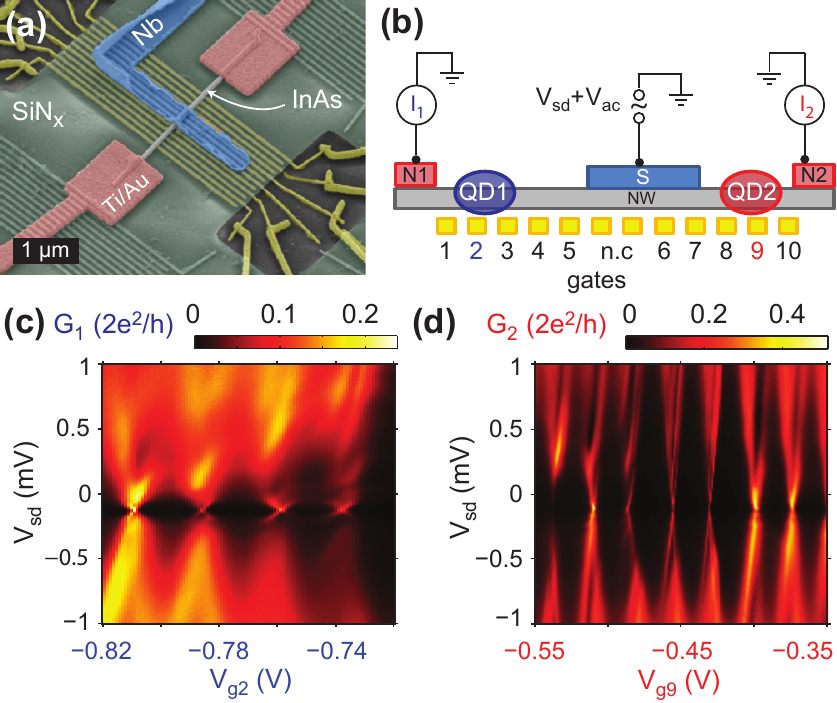}
\caption{(Color online) (a) SEM image of a representative CPS device. The InAs nanowire on the SiN$_x$ layer (green) is contacted by a central Nb (S, blue) and two Ti/Au leads (N1 and N2, purple). The local gates below the SiN$_x$ are colored yellow. (b) Schematic of the device and the measurement setup. Gates 1-3 are used to form QD1 and gates 8-10 for QD2. Gates 6 and 7 are below S. Two more gates between g5 and g6 below S were not connected (nc) and left floating in the experiments. (c) and (d) Differential conductance of QD1 and QD2 as a function of the bias, $V_{\rm SD}$, and the respective local tuning gates, $V_{\rm g2}$ and $V_{\rm g9}$.}
\label{Figure1}
\end{figure}

An artificially colored SEM image of a sample is shown on Fig.~\ref{Figure1}a. First, using electron-beam lithography, an array of twelve local gates was fabricated on a highly doped silicon substrate that serves as a global backgate, insulated by $\sim400\,$nm SiO$_2$. The local gates consist of $4\,$nm Ti and $18\,$nm Pt and are $\sim40\,$nm wide with an edge to edge separation of $\sim60\,$nm. These gates are overgrown by $\sim25\,$nm SiN$_x$ for electrical insulation using plasma-enhanced chemical vapor deposition. The SiN$_x$ was removed at the edges of the gate array by a reactive ion etch (RIE) with CHF$_3$/O$_2$\cite{Wong1992} to fabricate electrical contacts to the local gates. In the experiments only ten gates were used for technical reasons, so that two of the four gates below S were not connected (nc) and left floating. In the next step we deposit a single InAs NW ($\sim 70\,$nm diameter) perpendicular to the gates using micromanipulators. The NWs were grown by solid-source molecular beam epitaxy,\cite{Madsen_Nygard_JCrystGrowth_2013} implementing a two-step growth process to suppress stacking faults.\cite{Shtrikman_Heiblum_Nanolett_2009} The $330\,$nm wide and $40\,$nm thick superconducting Nb contact and the two normal metal drain electrodes ($7$/$95\,$nm Ti/Au) were fabricated in consecutive lithography steps, with prior ammonium sulfide passivation\cite{Suyatin2007} to remove the native oxide on the NW. 

The experiments were carried out in a dilution refrigerator with a base temperature $T\approx50\,$mK. As illustrated in Fig.~\ref{Figure1}b, we applied a sinusoidal voltage $V_{\rm ac}\approx 10\,\mu$V to the superconductor S and simultaneously recorded the resulting variations of the currents in the contacts N1 and N2, $I_1^{(\rm ac)}$ and $I_2^{(\rm ac)}$, using current-voltage (IV) converters and lock-in amplifiers. We define the differential conductances through QD $i$ as $G_i=I_i^{(\rm ac)}/V_{\rm ac}$. The lever arms of the different gates were found by bias spectroscopy experiments, applying a dc voltage to S. If not stated otherwise, all presented experiments were done at zero dc bias, which was achieved by compensating offsets in the IV-converters by external voltage sources (not shown in Fig.~\ref{Figure1}b).

The local gates g$i$ are numbered consecutively, as illustrated in Fig.~\ref{Figure1}b. The tunnel barriers for the QDs are formed for the conduction band electrons in the InAs NW by applying strongly negative voltages to the local gates. For QD1, g1 and g3 are used to induce the barriers (barrier gates) and g2 to tune the dot's chemical potential (tuning gate). QD2 is formed similarly with the gates g8 and g10, using g9 to tune the QD resonances. The other gates are kept on ground. The exact gate voltage settings in the presented experiments can be found in Tab.~\ref{tab:GateVoltages} in the appendix. In all figures we use the colors blue and red to distinguish the gates near QD1 and QD2, respecively. Fig.~\ref{Figure1}c shows the differential conductance $G_1$ as a function of the applied dc bias and $V_{\rm g2}$, from which we estimate a charging energy for QD1 of $1\,$meV. The lever arms obtained from similar experiments with the other gates suggest that QD1 is indeed formed between g1 and g3. At low bias some resonances occur, reminiscent of Andreev bound states,\cite{Pillet_NaturePhys6_2010, Schindele_Baumgartner_PRB89_2014} which suggests a relatively strong coupling to S and a weaker coupling to N1. From these states we deduce an effective superconducting energy gap on or near QD1 on the order of $\Delta^{*}\approx35\,\mu$eV. This gap considerably smaller than the bulk value of Nb ($\sim1.45\,$meV,\cite{Novotny_JLTP_1975}), possibly due to the strong suppression of the proximity induced gap expected for relatively thick semiconducting NWs.\cite{DasSarma2013} As shown in Fig.~\ref{Figure1}d, QD2 exhibits clear Coulomb blockade diamonds and a negligibly small energy gap ($<5\,\mu$eV). QD2 forms between g8 and g10, as expected, with a charging energy of $1.5\,$meV.

\section{Tuning of a drain tunnel barrier}

\begin{figure*}[t]
\centering
\includegraphics{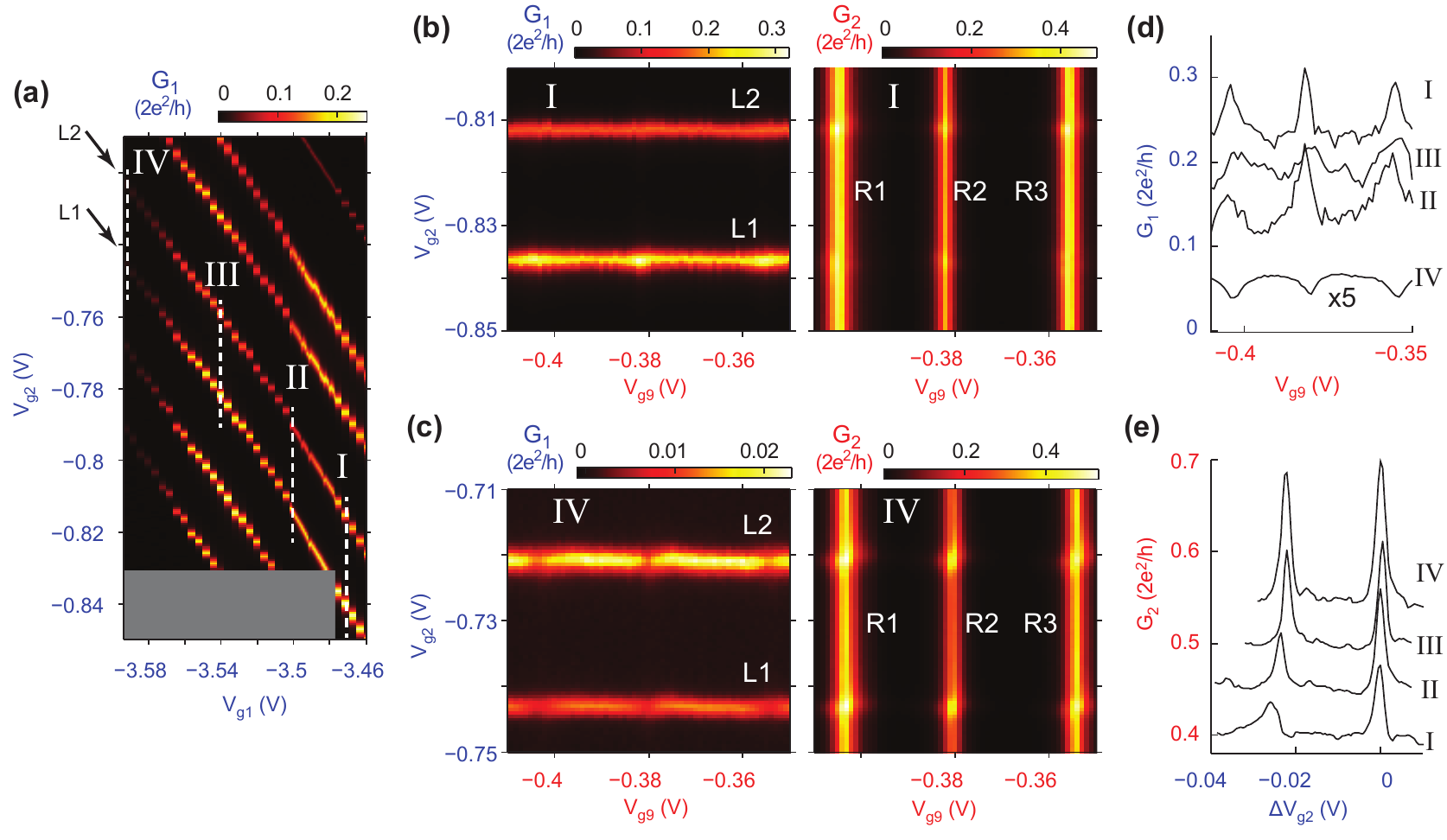}
\caption{(Color online) (a) $G_1$ as a function of the gate voltages $V_{\rm g1}$ and $V_{\rm g2}$, which shows the tuning of the tunnel coupling to the normal lead of QD1. The dashed lines indicate the settings for the following experiments. (b) $G_1$ and $G_2$ as a function of $V_{\rm g2}$ and $V_{\rm g9}$, measured in setting I of Fig.~a. (c) Conductance maps similar to (b) for the same QD states for setting IV in Fig.~a. (d) and (e) Evolution of the resonance maxima in $G_1$ and $G_2$ for the $V_{\rm g1}$-settings indicated in Fig.~a. The respective resonance crossings are labeled in brackets, e.g. (L1,R1). In (d) curve IV is multiplied by $5$ and in (e) all curves are offset vertically for clarity and centered to the L2 resonance.}
\label{Figure2}
\end{figure*}

Cooper pair splitting is a a nonlocal two-particle process and leads to a nonlocal signal, i.e., a signal that depends on the transmissions of both QDs. Experimentally, we investigate the change of conductance in one QD when the other dot is brought into resonance.\cite{Hofstetter2009, Schindele_Baumgartner_Schoenenberger_PRL109_2012} Competing processes, e.g., the sequential tunneling of Cooper pairs through the same QD (local pair tunneling), are local in nature and depend intrinsically only on the settings of one QD.
The aim of this work is to investigate the evolution of the nonlocal signal in a CPS device when one tunnel barrier of a QD is varied. In this section we tune the local gate g1 to change the tunnel coupling of QD1 to lead N1. Due to the close proximity of g1 to the center of QD1, this also changes the chemical potential of the dot, which we compensate using the local gate g2 (tuning gate). This procedure allows us to compare the same Coulomb blockade (CB) resonance for different tunnel barrier strengths. The differential conductance $G_1$ as a function of the two gates g1 and g2 is plotted in Fig.~\ref{Figure2}a. The reduction of the CB resonance widths and amplitudes suggests a variation of the involved tunnel barrier. Intuitively, g1 tunes $\Gamma_{\rm N1}$, i.e., the single electron tunnel coupling to N1. We expect that $\Gamma_{\rm N1}$ decreases when $V_{\rm g1}$ is made more negative, i.e., from $V_{\rm g1}$-position I in Fig.~\ref{Figure2}a to position IV. However, on a larger gate voltage scale the modulation of the resonance amplitude exhibits more than a single maximum, in contrast to what one might expect from tuning a simple tunnel barrier in a transport broadened QD. We attribute this experimental finding to the fact that the gates also tune other parts of the device, though by a considerably smaller lever arm.

To investigate the nonlocal signals in the CPS device, we simultaneously record $G_1$ and $G_2$ as a function of the tuning gates g2 and g9, as shown in Fig.~\ref{Figure2}b. While g2 tunes QD1 trough the two resonances L1 and L2 labeled in Fig.~\ref{Figure2}a, g9 tunes QD2 through the three resonances R1, R2 and R3. The resonances of the two QDs run perpendicular in these plots, which shows that the capacitive cross talk between the QDs is very small. Though not shown, we note that the conductance through QD1 and QD2 in series does not exhibit anti-crossings, which suggests that the inter-dot tunnel coupling is considerably smaller than the life-time broadening of the CB resonances.

The amplitude of one QD resonance is independent of the gate voltage applied to the other QD, except where both QDs become resonant with the Fermi energy in the leads. In Fig.~\ref{Figure2}b (configuration I) both conductances increase at these resonance crossings,\cite{footnote1} for which we use the term {\it positive correlation} between the conductance variations in the QDs. This positive correlation is characteristic for CPS,\cite{Schindele_Baumgartner_Schoenenberger_PRL109_2012} as we discuss in more detail below. Similar gate sweeps over the same resonances in configuration IV are plotted in Fig.~\ref{Figure2}c. While the QD2 resonances are similar as in setting I, the (local) conductance of QD1 is decreased by about an order of magnitude due to the increased barrier strength. Focusing on the resonance crossings one finds that the amplitudes of the QD1 resonances are {\it reduced} at the resonance crossings, while the QD2 resonances still exhibit an increased conductance, which results in a {\it negative} conductance correlation between the two QDs, in contrast to gate configuration I. The nonlocal signal on QD2 only changes in amplitude, but not in sign. We postpone the discussion of the origin of these dips to section VI and only point out that 1) the QD1 conductance away from the QD2 resonances is determined by the local processes and changes significantly between the gate configurations I-IV, as expected if the tunnel barrier strength is varied. 2) different neighboring QD states of similar amplitudes and widths can exhibit different conductance correlations (not shown), excluding electrostatic effects. Also resistive cross talk\cite{Hofstetter2009} can be excluded as the origin of the observed effects because it would lead to a dip in both conductances at a resonance crossing.

The evolution from a positive to a negative conductance correlation with the voltage on the local gate g1 can be followed better in Fig.~\ref{Figure2}d, where the amplitude of the QD1 resonance L1 is plotted as a function of the voltage applied to QD2-gate g9, $V_{\rm g9}$, for the four $V_{\rm g1}$-settings indicated in Fig.~\ref{Figure2}a. We observe three peaks where g9 tunes QD2 through the resonances R1-R3 and label each crossing by the two respective resonances, e.g. (L1,R1) for the gate configuration where L1 and R1 are both resonant. The conductance variation on these crossings are similar for the settings I and II, but decrease significantly for setting III. For configuration IV, we find a dip instead of a peak at the resonance crossings. For all four curves the local conductance background and the nonlocal conductance variations both decrease with more negative $V_{\rm g1}$. We note that no offsets are subtracted in Fig.~\ref{Figure2}d and curve IV is multiplied by $5$. The evolution of the nonlocal signal on QD1 has to be compared to the one on QD2: in Fig.~\ref{Figure2}e the amplitude of the QD2 resonance R1 is plotted for the same $V_{\rm g1}$-settings I-IV. Because the local conductance background is almost identical for all curves, II-IV are offset for clarity. For all four gate configurations we find a peak in the conductance as R1 crosses L1 and L2. With decreasing $V_{\rm g1}$, the conductance variation at the resonance crossings increases in amplitude by almost a factor of 2 between I and IV.

As a measure for the CPS efficiency we use $s=\frac{2G_{\rm CPS}}{G_1+G_2}$, which essentially compares the fraction of currents due to CPS to the total current in the system.\cite{Schindele_Baumgartner_Schoenenberger_PRL109_2012} If the conductance variations are the same in both QDs, one obtains a conservative estimate for $s$ by setting $\Delta G_{\rm CPS}=\Delta G_1=\Delta G_2$, with $\Delta G_i$ the conductance variations on the resonance crossings. This is applicable for the $V_{\rm g1}$-setting I, for which we find, for example for the resonance crossing (L1,R2), an efficiency of $s\approx$17\%. Clearly, we cannot use this approximation for the cases II-IV. To describe the transition from a positive to a negative correlation of the QD conductances one might also use the {\it visibility} of the nonlocal signal in one branch of the CPS device, which is defined as $\eta_i=\Delta G_i/G_i$ at a resonance crossing.\cite{Schindele_Baumgartner_Schoenenberger_PRL109_2012} For the resonance crossing (L1,R2) we find that $\eta_1$ decreases from $\sim20$\% to about $-40$\% when $V_{\rm g1}$ is tuned from configuration I to IV, while $\eta_2$ increases from $\sim8.5$\% to $\sim28.7$\%. The evolution of the visibilities directly illustrates the sign change of the conductance correlations between the two QDs when the tunnel coupling of QD1 to N1 is reduced.

\section{Tuning of a source tunnel barrier}

\begin{figure}[b]
\centering
\includegraphics{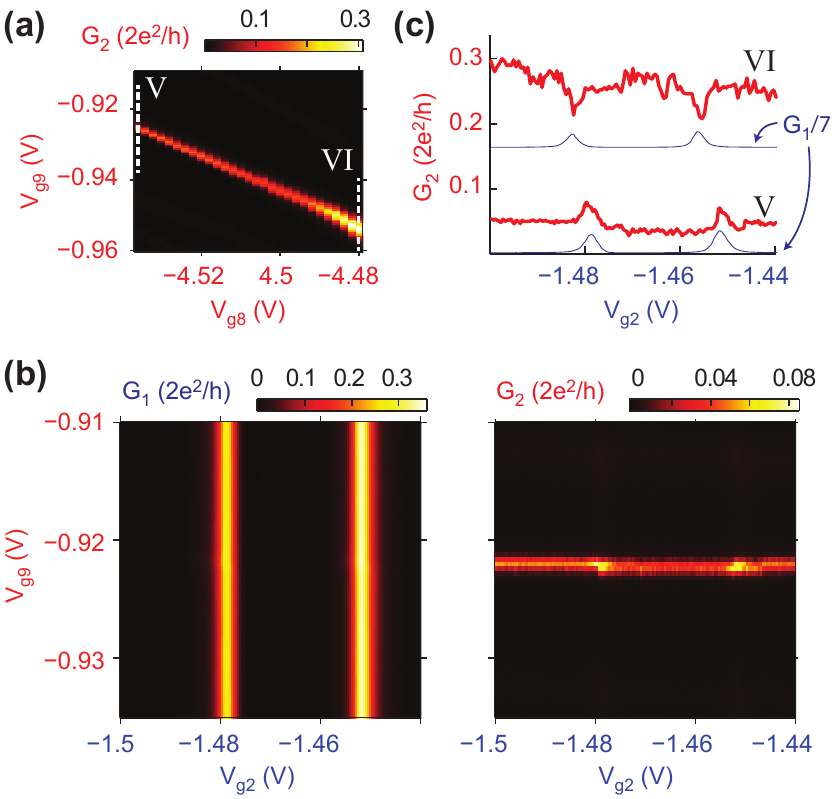}
\caption{(Color online) (a) Differential conductance of QD2, $G_2$ as a function of the two local gates g8 and g9. Two gate configurations, V and VI, are indicated by colored lines. (b) $G_1$ and $G_2$ as a function of the QD1-gate g2 and QD2-gate g9 showing two resonance crossings in the gate configuration V. (c) Amplitude of the QD2 resonance for g8-settings V and VI when QD1 is tuned through the two respective resonances shown in b. $G_1$ with a scale factor and an offset is plotted as a reference.}
\label{Figure3}
\end{figure}

In this section we investigate the evolution of the conductance correlations in the CPS device when tuning gate g8, which forms the barrier of QD2 to the superconductor S, see Fig.~\ref{Figure1}b. The exact gate voltages used to form the QDs are given in the appendix.\cite{footnote2} We show that a transition from a positive to a negative conductance correlation between the QDs can be induced by increasing the tunnel coupling of QD2 to S, similar as in Section III for a decreasing coupling of QD1 to N1. For simplicity, we only focus on a single resonance of QD2, whose differential conductance, $G_2$, is plotted in Fig.~\ref{Figure3}a as a function of the voltages applied to the QD2 gates g8 (barrier to S) and g9 (tuning gate of QD2). With a more negative voltage on g8, the resonance amplitude and width decrease markedly. Similarly as discussed for $\Gamma_{\rm N1}$ in the previous section, this probably corresponds to a stronger barrier and a weaker coupling $\Gamma_{\rm S2}$ to S. The two vertical lines labeled V and VI are the two $V_{\rm g8}$-settings for which we now investigate the nonlocal signals.

The conductances of QD1 and QD2, $G_1$ and $G_2$, are plotted in Fig.~\ref{Figure3}b as a function of $V_{\rm g2}$ (tuning gate of QD1) and $V_{\rm g9}$ (tuning gate of QD2) for $V_{\rm g8}$-setting V (see Fig.~\ref{Figure3}a). While g2 tunes QD1 through two resonances, g9 tunes QD2 through the resonance shown in Fig.~\ref{Figure3}a. Also here we do not find a significant capacitive or tunnel coupling between the QDs compared to the life time broadening. On the resonance crossings, we observe small peaks in $G_1$ (visibility $\eta_1\approx2.8$\%) and more pronounced peaks in $G_2$ ($\eta_2\approx48$\%), see Fig.~\ref{Figure3}b. Again we take the positive correlation between the nonlocal conductance variations as an indication for CPS. However, we could not tune these resonances to a $V_{\rm g8}$-setting for which $\Delta G_1=\Delta G_2$. The amplitude of the QD2 resonance as a function of the QD1 gate g2 is plotted in Fig.~\ref{Figure3}c for the gate configurations V and VI with the scaled QD1 resonances for orientation. For configuration V with a weaker tunnel coupling of QD2 to S, we find an increase in $G_2$ at the resonance crossings, while it is reduced at the crossings in configuration VI, in which QD2 couples stronger to S. Because the nonlocal signals on $G_1$ are all positive (not shown), this corresponds to a transition from a positive to a negative conductance correlation with increasing coupling to S. A qualitatively similar sign change of the conductance correlation was found in section III when decreasing the tunnel coupling of QD1 to the normal lead N1.

\section{Tuning of the inter-dot coupling}

\begin{figure}[b]
\centering
\includegraphics{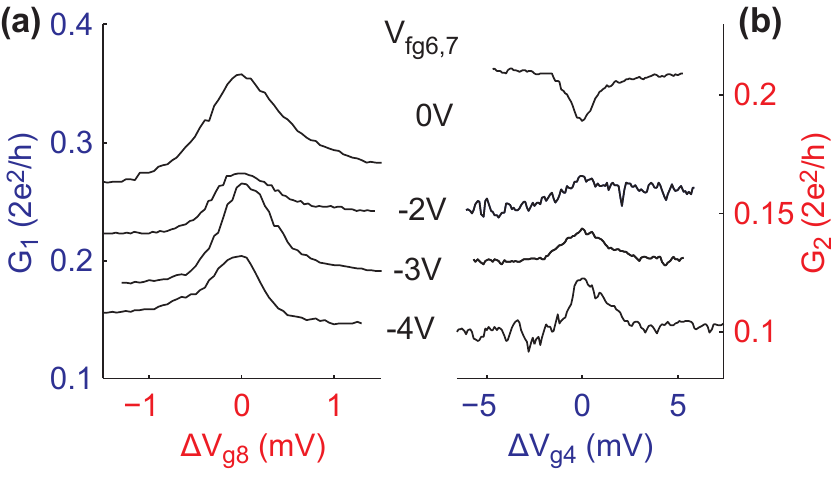}
\caption{(Color online) (a) $G_1$ as a function of the voltage on gate g8 ($V_{\rm g8}$), which tunes QD2 though a resonance for a series of voltages $V_{\rm g6}=V_{\rm g7}$ applied to the gates below S. The curves have the same background conductance and are shifted vertically for clarity and horizontally such that the resonance crossings occur at $\Delta V_{g8}=0$. (b) $G_2$ as a function of $V_{\rm g4}$, which tunes QD1 though a resonance. The curves are shifted similarly as the ones in (a).}
\label{Figure4}
\end{figure}

In a third experiment we defined two QDs closer to the superconductor using only two gates, namely g4 and g5 for QD1 and g8 and g9 for QD2. The exact gate voltages are given in the appendix. We use barrier gate g8 to also tune the chemical potential of QD2, and similarly gate g4 to tune QD1. The aim is to investigate the effect of the gates g6 and g7 {\it below} the superconducting contact S on the conductance correlations in the CPS device. Because of the finite size of the NW and despite the screening by the superconductor, we expect that the electron density below S is reduced when the gates are set to more negative potentials, which should lead to a reduction of the single electron tunneling rate between the QDs. In Fig.~\ref{Figure4}a the amplitude of a QD1 resonance ($G_1$) is plotted as a function of $V_{\rm g8}$, i.e., the gate defining QD2, for a series of voltages applied to the gates below S. In all experiments we set $V_{\rm g6}=V_{\rm g7}$. For the same voltages on $V_{\rm g6}$ and $V_{\rm g7}$ the amplitude of a QD2 resonance ($G_2$) is plotted in Fig.~\ref{Figure4}b as a function of $V_{\rm g4}$, i.e. a gate of QD1. The respective four curves in Figs.~\ref{Figure4}a and \ref{Figure4}b have the same (local) background  conductance within experimental error and are shifted vertically for clarity. This suggests that the tunnel barriers to the source and drain contacts are not significantly altered by the gates g6 and g7. In addition, the curves are shifted horizontally, so that the resonance crossings are centered at $\Delta V_{\rm g8}=0$ and $\Delta V_{\rm g4}=0$, respectively. This is necessary because these gates affect the resonance position of both QDs by a small capacitive coupling. For $V_{\rm g6}=V_{\rm g7}=0$ the nonlocal signal on QD1 is positive, but negative on QD2, so that we find a negative conductance correlation at the resonance crossing. When we continuously tune both gate voltages to more negative values, the nonlocal signal on QD2 at the resonance crossing evolves from negative to positive values, while the signal on QD1 is only slightly reduced. At $V_{\rm g6}=V_{\rm g7}=-4\,$V we find a positive correlation and similar amplitudes for the QD conductances. The visibilities in the two arms evolve with decreasing voltage from $\eta_1=42$\% and $\eta_2=-23$\% at $V_{\rm g6}=V_{\rm g7}=0$ to $\eta_1=26$\% and $\eta_2=17$\% at $V_{\rm g6}=V_{\rm g7}=-4\,$V.

\section{Rate equation model}
\label{sec:model}

In the experiments presented above we find large qualitative and quantitative differences in the conductance variations for different crossings of resonances of the two QDs. These `nonlocal' signals are surprisingly simple to tune from a positive to a negative correlation. In fact, we can induce such transitions by using any single local barrier gate. In this section we present a simple toy model (similar to the one in Ref.~\citenum{Schindele_Baumgartner_Schoenenberger_PRL109_2012}), which qualitatively describes the experimental findings and allows to identify the physical mechanisms that could lead to the observed transitions in the conductance correlations. A more involved model can be found, for example, in Ref.~\citenum{Chevallier_Martin_PRB83_2011}. The two basic ideas are that 1) both, the local and nonlocal processes depend on the QD occupations, which couples the resulting rates, and 2) a finite inter-dot coupling can lead to electrons tunneling between the QDs, i.e., a local process that depends on the occupation of both QDs.

\begin{figure}[t]
\centering
\includegraphics{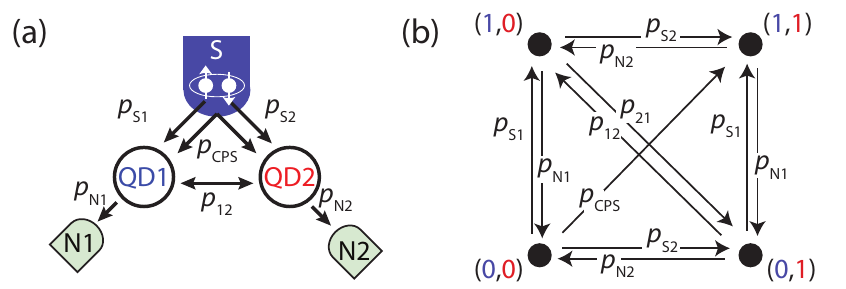}
\caption{(Color online) (a) Schematic of the device and transition probabilities. (b) Schematics of transitions between the allowed system states.}
\label{Figure5}
\end{figure}

\begin{table} [b]
	\centering
		\begin{tabular}{|c|c|c|c|} \hline
						Process & diagram & rate & transitions\\ \hline \hline
						
			SET to N1&
\includegraphics[width=0.15\textwidth]{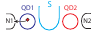}			
			 & $\Gamma_{\rm N1}$ &
			 \parbox{2cm}{$(1,0)\rightarrow(0,0)$ \\ $(1,1)\rightarrow(0,1)$} \\ \hline
			 
			\parbox{2cm}{LPT \\ into lead N$_{1}$} & 	\includegraphics[width=0.15\textwidth]{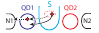}	 & $\Gamma_{S1}^2\Gamma_{N_1}$ &
			\parbox{2cm}{$(0,0)\rightarrow(1,0)$ \\ $(0,1)\rightarrow(1,1)$}\\ \hline
			
						CPS & 	\includegraphics[width=0.15\textwidth]{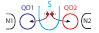}	 & $\Gamma_{S1}\Gamma_{S_2}$ &
						$(0,0)\rightarrow(1,1)$\\ \hline
			
						\parbox{2cm}{SET \\ between QDs}& 	\includegraphics[width=0.15\textwidth]{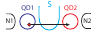}	 & $\Gamma_{12}$ &
						\parbox{2cm}{$(1,0)\rightarrow(0,1)$ \\ $(0,1)\rightarrow(1,0)$}\\ \hline
			
						\parbox{2cm}{SCPS \\ via QD $1$}& 	\includegraphics[width=0.15\textwidth]{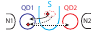}	 & $\Gamma_{S1}^2\Gamma_{12}$ & 
						$(0,0)\rightarrow(1,1)$\\ \hline
						
			\parbox{2cm}{SET from S\\ to QD1}&
	\includegraphics[width=0.15\textwidth]{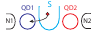}			
			 & $\Gamma_{\rm S1}$ &
			 \parbox{2cm}{$(0,0)\rightarrow(1,0)$ \\ $(0,1)\rightarrow(1,1)$} \\ \hline
			
		\end{tabular}
	\caption{Single electron and Cooper pair transport processes taken into account in the model: acronym, rate and transition in QD occupancies.}
	\label{tab:ProcessesCPS}
\end{table}

We first describe the model in some detail. Because of the large charging energy, each QD can only be empty, or occupied by a single electron at a time, i.e., the system occupies the states (0,0), (1,0), (0,1) or (1,1), which correspond to no electrons in the system, one in QD1 or in QD2, or an electron in both QDs, respectively. As illustrated in Fig.~\ref{Figure5}, we consider several processes that lead to transitions between the system states. The respective rates are determined by the tunnel couplings of the QDs to the three contacts and the inter-dot coupling. We use the classical, intuitive expressions for these rates, as listed in Tab.~\ref{tab:ProcessesCPS}, where $\Gamma_{N1}$ and $\Gamma_{N2}$ are the couplings to the normal metal contacts, $\Gamma_{S1}$ and $\Gamma_{S2}$ to S, and $\Gamma_{12}$ is the direct coupling between the QDs. The steady state QD occupations of the states (i,j), $P_{(i,j)}$, were calculated from a set of classical rate equations, using a diagrammatic method.\cite{Schankenberg_RevModPhys_1976} The processes we consider here are 1) tunneling of an electron from a filled QD $i$ to the respective normal electrode with the rates $\Gamma_{Ni}$ (denoted by SET, single electron tunneling). This process leads to a current in the respective contact. 2) local pair tunneling (LPT), where the electrons of a Cooper pair (CP) are transmitted sequentially through the same QD $i$. This requires the QD to be empty initially and leads to an electron emitted to lead $N_i$ and to the occupation of QD $i$ by the second electron. The probability of this process scales with $\Gamma_{\rm Si}^2\Gamma_{\rm Ni}$. 3) Cooper pair splitting, where the electrons of a Cooper pair tunnel into two initially empty QDs. CPS scales as $\Gamma_{\rm S1}\Gamma_{\rm S2}$ and leads to two full QDs, but not directly to a current in the normal leads. 4) Here we also investigate in more detail the effect of a single electron tunnel coupling between the QDs (SET between QDs), which scales directly with $\Gamma_{12}$. We note that the inter-dot coupling is not necessarily a direct single electron process, but could also be due to higher order processes mediated by the superconductor. 5) Because of a possibly large inter-dot coupling, we also consider processes in which Cooper pair electrons sequentially tunnel to one QD and the first leaves the dot by tunneling to the other QD. We call these processes {\it sequential CPS} (SCPS), stressing that they lead to a transition of two empty dots to two filled dots, similar to direct CPS. Sequential CPS scales as $\Gamma_{\rm Si}^2\Gamma_{12}$ and do not lead directly to a current in the normal leads. 6) As a last process we also consider the tunneling of a single electron from S to one of the QDs, which scales with $\Gamma_{Si}$. This process should be suppressed  for a superconductor with an ideal energy gap at zero temperature. In addition, we generally assume that electrons effectively tunnel only in the direction from S to N1 or N2.

Each process should be weighted in addition with individual prefactors accounting for the density of states, differing effects of the superconductor's energy gap (e.g., a 'soft gap' due to the breaking of Cooper pairs at material interfaces, which allows the injection of single electrons) and the inverse scaling of the CPS probability with the separation between the emission positions of the two Cooper pair electrons.\cite{Recher_Loss_PRB63_2001} Since we only aim for a qualitative picture we simply set the prefactors for SET from S and for CPS to $k=0.1$ and all other prefactors to 1. We note that these prefactors are crucial for a quantitative determination of the CPS efficiency, which is beyond the scope of this simple model, and we use a fixed resonance width, independent of the tunnel couplings.

\begin{figure}[b]
\centering
\includegraphics{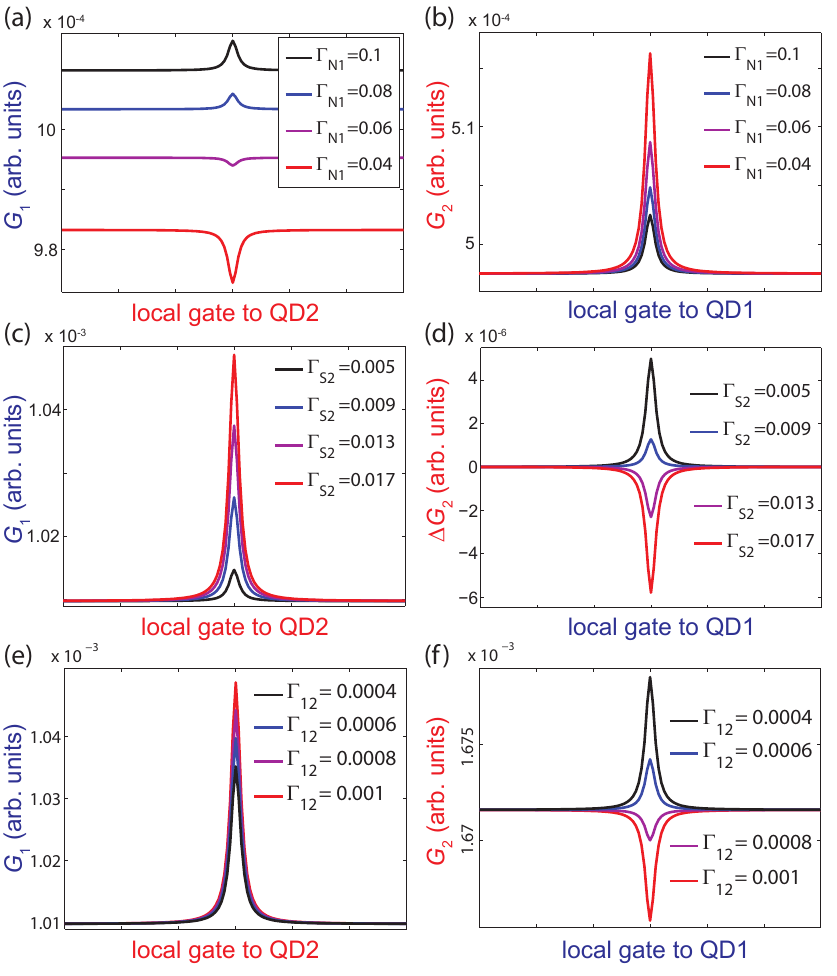}
\caption{(Color online) Results of the model calculation showing the transition between positive and negative correlations between the conductance variations on a resonance crossing. If not stated otherwise in the subfigures, the tunnel couplings are set to \textcolor{blue}{$\Gamma_{S1}=0.01$, $\Gamma_{N1}=0.1$}, 
\textcolor{red}{$\Gamma_{S2}=0.005$, $\Gamma_{N2}=0.05$} and $\Gamma_{12}=0.001$. (a) and (b) Transition induced by tuning $\Gamma_{N1}$, (c) and (d) transition induced by tuning $\Gamma_{S2}$, and (e) and (f) transition induced by tuning $\Gamma_{12}$, for fixed to  $\Gamma_{S2}=0.017$.}
\label{Figure6}
\end{figure}

We calculate the conductance into N1 (similar for N2) from the average system state occupation, $P_{(i,j)}$, and the rates for local SET and LPT to N1:
$$ G_{N1}/G_0=\Gamma_{N1}\left[P_{(1,0)}+P_{(1,1)}\right]+ \Gamma_{S1}^2\Gamma_{N1}\left[P_{(0,1)}+P_{(0,0)}\right],$$
with $G_0$ the conductance quantum.

Similar expressions can easily be derived for the other conductances in the system. In particular, the contribution of CPS can be found as
$$ G_{\rm CPS}/G_0=k\Gamma_{S1}\Gamma_{S2} P_{(0,0)}.$$

A first important finding in this model is that positive conductance correlations between the two QDs only if the CPS rate is non-zero. In other words: even with many other processes involved, CPS can be identified qualitatively by a positive correlation of the nonlocal signals. A negative correlation between the nonlocal signals, however, can have different origins: A) with a finite inter-dot coupling the current through one QD can be partially diverted to the other QD, thereby decreasing the current to one normal contact and increasing the current to the other. In this scenario no nonlocal processes are required to obtain a negative conductance correlation. B) On each QD the local processes and CPS compete for the dot occupation. For example, switching on CPS by bringing QD2 into resonance leads to an increase of the average QD1 occupation, which reduces the frequency of the local processes. For this mechanism no inter-dot coupling is required.

In Fig.~\ref{Figure6} the conductances obtained in this model through QD1 (left column) and QD2 (right column) into the respective normal metal contacts are plotted for a series of systematic changes of a single tunnel coupling, with all other parameters held constant (the values are given in the caption to Fig.~\ref{Figure6}). Figs.~\ref{Figure6}a and Fig.~\ref{Figure6}b show the evolution from a positive to a negative conductance correlation when reducing $\Gamma_{N1}$, similar to the experiments in Section III. In our model we can trace this transition to an increased population of QD1 when the barrier to N1 is made more opaque, so that the tunnel coupling to QD2 becomes more relevant as a path to emit electrons from QD1. It is interesting to note that in our model we were not able to generate strong negative conductance correlations similar to the experiments without including quasi particle tunneling from S.

Figures~\ref{Figure6}c and \ref{Figure6}d show the effect of tuning the coupling of QD2 to the superconductor S. Here the increased coupling to S results in an increase of the QD2 population (`stronger filling rate') and a transition from a positive to a negative conductance correlation. In particular, a {\it weaker} barrier to S has a similar effect as a {\it stronger} barrier to the normal metal contact, in qualitative agreement with the experiments in sections III and IV.

In Figs.~\ref{Figure6}e and Fig.~\ref{Figure6}f the effect of tuning the inter-dot coupling is investigated, which should be compared to the experiments in Section V. Here we start with a negative conductance correlation by setting $\Gamma_{S2}=0.017$, i.e., QD2 has a relatively large average population (all other rates are the same as above). This leads to electrons traversing from QD2 to QD1 and therefore to a dip in $G_2$ and an increase in $G_1$. When the inter-dot coupling $\Gamma_{12}$ is reduced, the current from QD2 to QD1 is suppressed and we find a transition from the negative to the positive conductance correlation and a nonlocal signal determined mainly by CPS. We note that the relation of the inter-dot coupling to the gates g6 and g7 in the experiments in Section IV is quite intuitive, since they probably tune the electron density below S and thus might pinch-off the coupling between the QDs.

\section{Conclusions and Outlook}
In summary, we report the tuning of the nonlocal signals by local bottom gates in a Cooper pair splitter device with a Nb contact. We find strong systematic transitions between positive and negative conductance correlations on resonance crossings, which can be explained qualitatively by the electron dynamics on the double dot system and Cooper pair splitting. In the presented simulations it is clear that the CPS part is modulated strongly by tuning the local gates. However, in the experiments the different contributions to the conductances are difficult to disentangle. The recovery of the positive correlations with all relevant gates strongly suggests that the CPS signal can be optimized using local gating techniques, which is an important step towards a reproducibly working source of entangled electron pairs.

\begin{acknowledgments}
We thank the group of M. Poggio for the support with the micromanipularors and B. Braunecker, F. Domingues, A. Levi Yeyati, P. Moca and G. Zarand for fruitful discussions. We gratefully acknowledge the financial support by the EU FP7 project SE$^2$ND, the EU ERC projects CooPairEnt and QUEST, the Swiss NCCR QSIT, the Swiss SNF, and the Danish Research Councils.
\end{acknowledgments}

\appendix

\section{Gate voltages to form the quantum dots}

Table~\ref{tab:GateVoltages} lists the voltages applied on the local gates to form the experiments presented in Figs.~1-4. The voltages defining the barriers of QD1 are given in blue, the ones defining the barriers of QD2 in red. Gates below S are colored in green. The backgate was set to zero potential in all experiments. The QDs in the last experiments were defined by only two gates near S, while the gates near N1 and N2 were set to large positive voltages to increase the coupling to the normal contacts. Gate voltages tuned during the experiments are labeled '(t)'.

\begin{table}[h!]
	\centering
\begin{tabular}{|c|c|c|c|}
\hline 
 $V_{\rm gi}$ (V) & Fig.~\ref{Figure1} and \ref{Figure2} & Fig.~\ref{Figure3} & Fig.~\ref{Figure4} \\ 
\hline 
\hline 
$\textcolor{blue}{V_{\rm g1}}$ & \textcolor{blue}{-3.5} & \textcolor{blue}{-3.475}  & +4 \\ 
\hline 
$\textcolor{blue}{V_{\rm g2}}$ & -0.8  (t) & -1.47  (t) & 0 \\ 
\hline 
$\textcolor{blue}{V_{\rm g3}}$ & \textcolor{blue}{-6.5}  & \textcolor{blue}{-6.5}  & 0 \\ 
\hline 
$\textcolor{blue}{V_{\rm g4}}$ & 0  & 0  & \textcolor{blue}{-4.13} \\ 
\hline 
$\textcolor{blue}{V_{\rm g5}}$ & 0  & 0  & \textcolor{blue}{-4.3} \\ 
\hline 
$\textcolor{green}{V_{\rm g6}}$ & 0  & 0  & (t) \\ 
\hline 
$\textcolor{green}{V_{\rm g7}}$ & 0  & 0  & (t) \\ 
\hline 
$\textcolor{red}{V_{\rm g8}}$ & \textcolor{red}{-4.5}  & \textcolor{red}{-4.5} & \textcolor{red}{-4.33} (t) \\ 
\hline 
$\textcolor{red}{V_{\rm g9}}$ & -0.38 (t) & -0.9  (t) & \textcolor{red}{-4.53} \\ 
\hline 
$\textcolor{red}{V_{\rm g10}}$ & \textcolor{red}{-4.7}  & \textcolor{red}{-4.7}  & +4 \\ 
\hline 
\end{tabular}

	\caption{Gate configurations in the different experiments.}

	\label{tab:GateVoltages}

\end{table}

\bibliographystyle{apsrev}

\begin{thebibliography}{99}
\bibitem{footnote1}
\bibinfo{note}{For resonance L2 this is difficult to discriminate because of the conductance scale.}
\bibitem{footnote2}
\bibinfo{note}{Due to a charge rearrangement we were not able to exactly identify the same resonances as used in section III.}
\end{thebibliography}

\end{document}